\documentclass{article}

\usepackage{arxiv}

\usepackage[utf8]{inputenc} 
\usepackage[T1]{fontenc}    
\usepackage{hyperref}       %
\usepackage{algorithm}
\usepackage{amssymb}
\usepackage{cite}
\usepackage{amsmath,amssymb,amsfonts}
\usepackage{url}            
\usepackage{booktabs}       
\usepackage{amsfonts}       
\usepackage{nicefrac}       
\usepackage{microtype}      
\usepackage{lipsum}
\usepackage{graphicx}
\graphicspath{ {./images/} }
\usepackage{pdfpages}
\usepackage{graphicx}
\usepackage{textcomp}
\newcommand*\circled[1]{\textcircled{\scriptsize #1}}

\usepackage{graphicx} \usepackage{hyperref} 
\makeatletter \def\orcidicon{\includegraphics[width=1em]{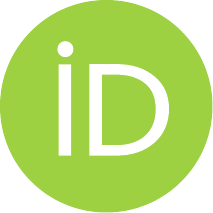}} 
\newcommand{\orcidlink}[1]{\href{https://orcid.org/#1}{\orcidicon}} \makeatother
\usepackage{arydshln}
\usepackage{color, colortbl}
\definecolor{LightBlue}{rgb}{0.81, 0.91, 0.98}
\definecolor{LightRed}{rgb}{0.98, 0.82, 0.83}
\usepackage{array} 
\usepackage{url}
\usepackage{algpseudocode}
\usepackage{booktabs}
\hypersetup{colorlinks=true,urlcolor=blue,linkcolor=black,citecolor=blue,urlcolor=blue,allcolors=blue }
\usepackage{float}
\usepackage{multirow} 
\usepackage{mathtools}
\usepackage{soul} 
\usepackage{listings}
\usepackage{color}

\definecolor{gray}{rgb}{0.5,0.5,0.5}
\definecolor{mauve}{rgb}{0.58,0,0.82}
\definecolor{dkgreen}{rgb}{0.0, 0.5, 0.0}
\title{MalVis: A Large-Scale Image-Based Framework and Dataset for Advancing Android Malware Classification}

\author{
 Saleh J. Makkawy~\orcidlink{0009-0006-3437-335X} \\
  Department of Electrical and Computer Engineering\\
  University of Delaware\\
  Newark, DE 19711 \\
  \texttt{salehmak@udel.edu} \\
   \And
Michael J. De Lucia \\
  Department of Electrical and Computer Engineering\\
  University of Delaware\\
  Newark, DE 19711 \\
  \texttt{mdelucia@udel.edu} \\
  \And
Kenneth E. Barner~\orcidlink{0000-0002-0936-7840}\\
  Department of Electrical and Computer Engineering\\
  University of Delaware\\
  Newark, DE 19711 \\
  \texttt{barner@udel.edu} 
}

\begin{document}
\maketitle
\begin{abstract}
As technology advances, developers continually create innovative solutions to enhance smartphone security. However, the rapid spread of Android malware poses significant threats to devices and sensitive data. The Android Operating System (OS) 's open-source nature and Software Development Kit (SDK) availability mainly contribute to this alarming growth. Conventional malware detection methods, such as signature-based, static, and dynamic analysis, face challenges in detecting obfuscated techniques such as encryption, packing, and compression in malware. Although developers have created several visualization techniques for malware detection using deep learning (DL), they often fail to identify the critical malicious features of malware accurately. This research introduces MalVis, a unified visualization framework that integrates entropy and N-gram analysis to emphasize meaningful structural and anomalous operational patterns within the malware bytecode. By addressing significant limitations of existing visualization methods, such as insufficient feature representation, limited interpretability, small dataset sizes, and restricted data access, MalVis delivers enhanced detection capabilities, particularly for obfuscated and previously unseen (zero-day) malware. The framework leverages the MalVis dataset introduced in this work, a publicly available large-scale dataset comprising more than 1.3 million visual representations in nine malware classes and one benign class. A comprehensive comparative evaluation was performed against existing state-of-the-art visualization techniques using leading convolutional neural network (CNN) architectures, MobileNet-V2, DenseNet201, ResNet50, and Inception-V3. To further boost classification performance and mitigate overfitting, the outputs of these models were combined using eight distinct ensemble strategies. To reduce the problem of an imbalanced class distribution in the multiclass dataset, we implemented an undersampling technique to ensure balanced learning across all types of malware. MalVis achieved superior results, with 95.19\% accuracy, 90.81\% F1-score, 92.58\% precision, 89.10\% recall, 87.58\% Matthews Correlation Coefficient (MCC), and 98.06\% Receiver Operating Characteristic Area Under Curve (ROC-AUC). These findings emphasize the effectiveness of MalVis in providing interpretable, accurate representation features for malware detection and classification, offering a valuable resource for both research and real-world security applications.
\end{abstract}


\section{Introduction}\label{sec:introduction}
{Smartphones} are proliferating, with projections indicating that they will exceed 7 billion by 2025, and nearly 70\% using the Android operating system~\cite{ahmedSherif2024}~\cite{statista2024}. Due to their compact designs, these mobile devices have become indispensable, facilitating tasks like email management, banking transactions, and the storage of sensitive health information. However, the widespread adoption of smartphones has also drawn the attention of hackers~\cite{MSI_2023}, exacerbated by the open-source nature of Android's OS and its SDK. This vulnerability has facilitated the way for various forms of malware, including viruses~\cite{noever2021virus},~\cite{wang2009understanding}, worms~\cite{kienzle2003recent}, adware~\cite{yilmaz2015adware,suresh2019analysis}, spyware~\cite{boldt2004exploring}, ransomware~\cite{ali2017ransomware}, rootkits~\cite{beegle2007rootkits}, trojans~\cite{zhenfang2015study}, keyloggers~\cite{bhardwaj2020keyloggers}, botnets~\cite{feily2009survey}, and mobileware~\cite{pachhala2021comprehensive}. Consequently, developing a robust defense system capable of identifying and mitigating this wide range of threats is crucial. While traditional detection methods like signature-based~\cite{halevi2006strengthening,canfora2015obfuscation}, dynamic analysis~\cite{DATDroid}, and static analysis~\cite{pan2020systematic} remain dominant, they often struggle with modern evasion techniques such as code obfuscation, encryption, polymorphism, and packing. As a result, there has been a growing interest in using advanced Deep Learning (DL) techniques to analyze malware and detect these suspicious behaviors.

S everal studies have explored the transformation of binary code or bytecode into image representations to leverage the capabilities of Deep Neural Network (DNN) for malware detection. However, these approaches frequently fail to capture semantic context, structural anomalies, and obfuscation pattern features critical for accurate and robust classification.

Designing effective detection systems, particularly those utilizing visual representations, requires a thorough understanding of the structural composition of Android applications. The following section presents an overview of the Android Package Kit (APK) file structure, emphasizing its core components relevant to malware analysis.

\subsection{Overview of the Android APK file structure}\label{background}
\begin{figure}[h]
  \centering
  \includegraphics[width=130mm]{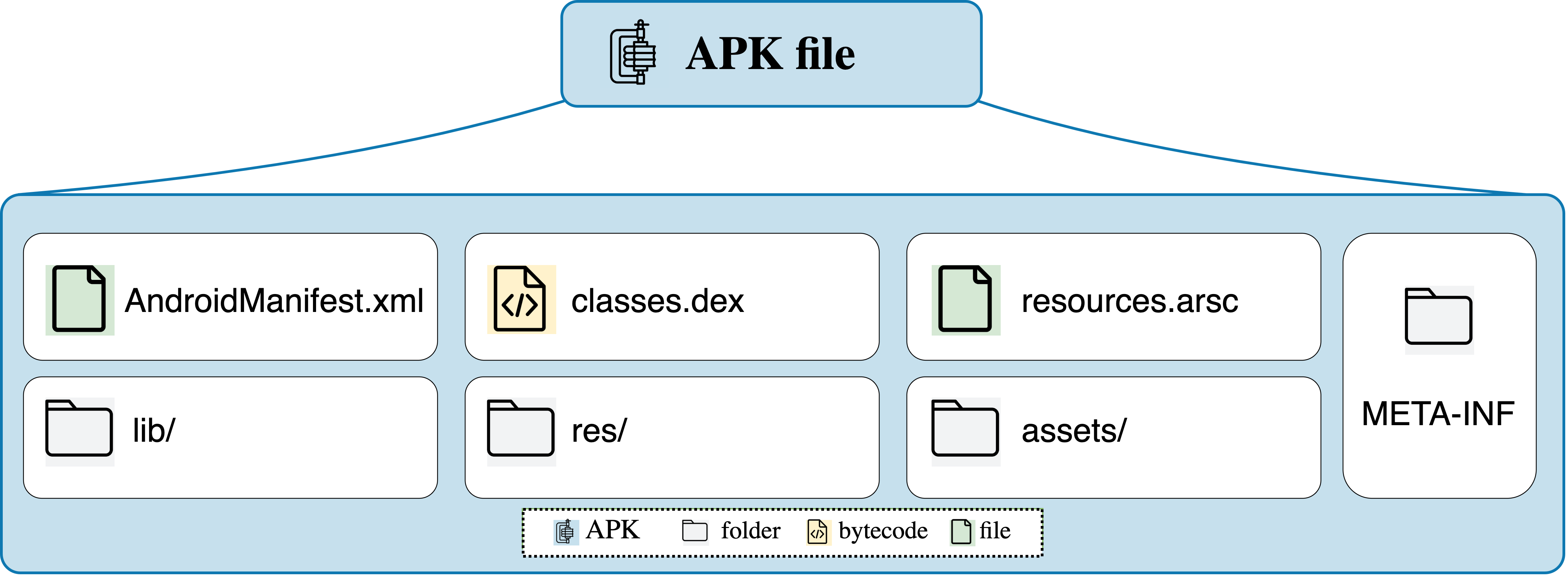}
  \caption{An illustration of the structure of an Android APK file, highlighting key components such as application bytecode, assets, resources, and the manifest file.}
  
  \label{fig: APK file structure}
\end{figure}

The APK is a compressed file that the Android OS uses to distribute and install applications, consisting of core files and folders such as the application bytecode, assets, resources, and a manifest file, as presented in Fig.~\ref{fig: APK file structure}.

Visualization approaches mainly focus on analyzing the \textit{AndroidManifest.xml} and \textit{Classes.dex} files, as these components provide critical insight into the structure and behavior of Android applications. The \textit{AndroidManifest.xml} file specifies essential metadata, including the application's components, package name, and requested permissions. In contrast, the \textit{Classes.dex} file contains the executable bytecode intended for the Android Dalvik Virtual Machine (DVM), making it a vital source for behavioral analysis. This research focuses on encoding and analyzing the \textit{Classes.dex} file, given its importance in capturing malicious behavior. Our proposed approach employs CNN models to detect hidden Android malware threats by transforming bytecode into visual representations, focusing on highlighting anomalous operational or structural patterns indicative of obfuscation or malicious intent.

\subsection{signature-based Analysis}
The signature-based detection method is widely used to recognize and detect malware. A software signature is a unique identifier that cannot be replicated and is typically generated using hash algorithms such as RSA, MD5, SHA1, SHA-256, and SHA-512 ~\cite{halevi2006strengthening,7346730}. The detection engine generates a signature for the software and compares it with a blocklist database of signatures stored locally or in the cloud. Typically, these blacklisted databases are proprietary assets of vendors, with restricted access granted to licensed users. A significant limitation of this method is the need for the detection engine to continuously refresh its blocklist database, which can lead to potential gaps in identifying new zero-day malware~\cite{poettering2022sequential}. As technology evolves, malware creators continue to find new techniques to evade detection, such as code alteration, function modification, file repacking, data encoding, or null byte injection, all to generate new signatures capable of evading security defenses ~\cite{canfora2015obfuscation,tahir2018study,gao2023obfuscation,elsersy2022rise,kirat2015malgene}.

\subsection{dynamic analysis}
Dynamic analysis is a pivotal method for malware detection, which involves observing and understanding software behavior during execution within a controlled and contained environment, such as a Sandbox or Virtual Machine (VM). This technique is effective in detecting abnormal actions, such as invoking suspicious system calls~\cite{pereberina2022approach}, examining network traffic~\cite{ullah2024nmal}, altering memory~\cite{sihwail2019malware}, and detecting errors in Logcat that invoke suspicious services from the OS~\cite{seyfari2023new}. 

However, this method requires accessing or monitoring users' sensitive information, which can be impractical when managing highly confidential data~\cite{orlova2021potential}. Despite its promising outcomes, acquiring an extensive dataset of labeled training data for optimal performance is often both time-consuming and costly~\cite{bhatia2017malware}. Security researchers are increasingly shifting to the visualization of malware based on static analysis images, which allows instant scanning of malware images to overcome the challenges posed by new malware~\cite{jose2020integrated,freitas2020large,vu2020hit4mal}. Unlike dynamic analysis methods that require days or weeks to monitor suspicious behavior in an application, these image representations can be harmless, do not require manual feature engineering, and resist typical obfuscation techniques employed by adversaries~\cite{sutter2024dynamic}.

\subsection{static analysis} 
Static analysis is a technique used to evaluate applications without executing them or observing their execution behavior, which can often be demanding and time-consuming. By not requiring execution, static analysis provides a unique assessment mode comparable to behavioral analysis techniques. One of the primary advantages of this malware detection method is its cost-effectiveness, as it minimizes the need for additional hardware or extensive computational resources beyond the actual analysis tool itself~\cite{pan2020systematic}. 

Despite its advantages, this approach has notable limitations. Specifically, it largely depends on identifying already known malware patterns, which challenges its effectiveness in generalizing and detecting evolving zero-day malware. Research efforts focus on improving the detection of suspicious activities using advanced methodologies such as machine learning (ML) and convolutional neural network (CNN)~\cite{kancherla2013image, makkawy2024improving} to mitigate this limitation. These innovations aim to improve the adaptability and robustness of static analysis against evolving threats.

\subsection{Contributions}
Our novel Android malware visualization framework uniquely integrates critical semantic and structural features extracted from executable bytecode and transforms them into RGB representations. Unlike previous techniques, MalVis enhances interpretability and classification accuracy while maintaining resilience against obfuscation.

Our contributions include the following:
\begin{itemize}
    \item \textbf{MalVis Dataset:}  Introducing MalVis, the largest Android malware visualization dataset with over 1.3 million images across ten classes, including nine malware types and benign software that is accessible to the research community\footnote{\href{www.mal-vis.org}{https://www.mal-vis.org}}. Scripts for generating these various visualization methods are publicly available on GitHub at the link\footnote{\href{https://github.com/makkawysaleh/MalVis}{https://github.com/makkawysaleh/MalVis}}.

    \item \textbf{Enhanced Visualization Framework:} Developing an advanced MalVis framework that enhances malware visualization by incorporating an entropy encoder with an N-gram technique. This approach utilizes the three RGB channels to effectively capture a broader range of malware characteristics, including encryption, compression, packing, and structural irregularities. This improves the precision of malware pattern detection in the visualizations.

    \item \textbf{Enhanced Multiclass Labeling:}
    Implementing an improved multiclass labeling approach using results from Euphony~\cite{hurier2017euphony} and VirusTotal~\cite{virustotal} allows precise classification and analysis of malware behavior, enhancing targeted threat identification and classification.

    \item \textbf{Robust Detection Model:}
    Evaluation of the performance of the MalVis framework on several SOTA visualization methods using advanced deep CNN architectures such as MobileNet-V2, DenseNet201, ResNet50, and Inception-V3, combined with several ensemble techniques, to further improve detection accuracy and generalization. The results showed that the MalVis framework achieved superior performance compared to others.
    
\end{itemize}
This research builds on our previous work, ``Improving Android Malware Detection Using a Bytecode-to-Image Encoding Framework''~\cite{makkawy2024improving} to detect anomalous structural and malicious features in Android malware. Although traditional detection techniques such as signature-based, static, and dynamic analysis remain prevalent, visualization-based approaches have gained popularity due to their speed and ability to highlight malicious patterns. However, existing methods often rely on simplistic byte-to-color mappings based on byte location in the file, which overlook semantic features and abnormal structure traits of malware. Additionally, they struggle against obfuscation, encryption, and packing. MalVis offers a richer and more interpretable representation that enhances the classification's robustness and addresses existing methods' limitations.

This paper is organized as follows. Section~\ref{sec:introduction} introduces the background of Android malware threats and provides an overview of the Android APK file structure, followed by traditional detection approaches and a summary of our key contributions. Section~\ref{sec:related} reviews related work, including the motivation behind visual-based detection, limitations of existing malware image datasets, and prior grayscale and RGB encoding techniques. Section~\ref{sec:method} details our proposed MalVis framework, including the data generation process, bytecode-to-image transformation, and an in-depth analysis of entropy and N-gram features through two distinct visualization approaches. Section~\ref{sec:evaluation} defines the performance evaluation metrics used to assess the model's effectiveness. In Section~\ref{sec:results}, we present experimental results across binary and multiclass classification tasks, evaluate the impact of data balancing using undersampling, and demonstrate performance improvements through ensemble modeling. Finally, Section~\ref{sec:conclusion} concludes the paper and outlines future research directions.

\section{Related Works}\label{sec:related}
This section describes the motivation to adopt visual representations in malware analysis. Next, we review Android malware visualization datasets that benchmark image-based malware detection. Finally, we discuss recent visualization-based detection techniques, focusing on grayscale and RGB encoding methods to detect malicious patterns.

\subsection{Motivation}
\begin{figure}[h]
  \centering
  \includegraphics[width=130mm]{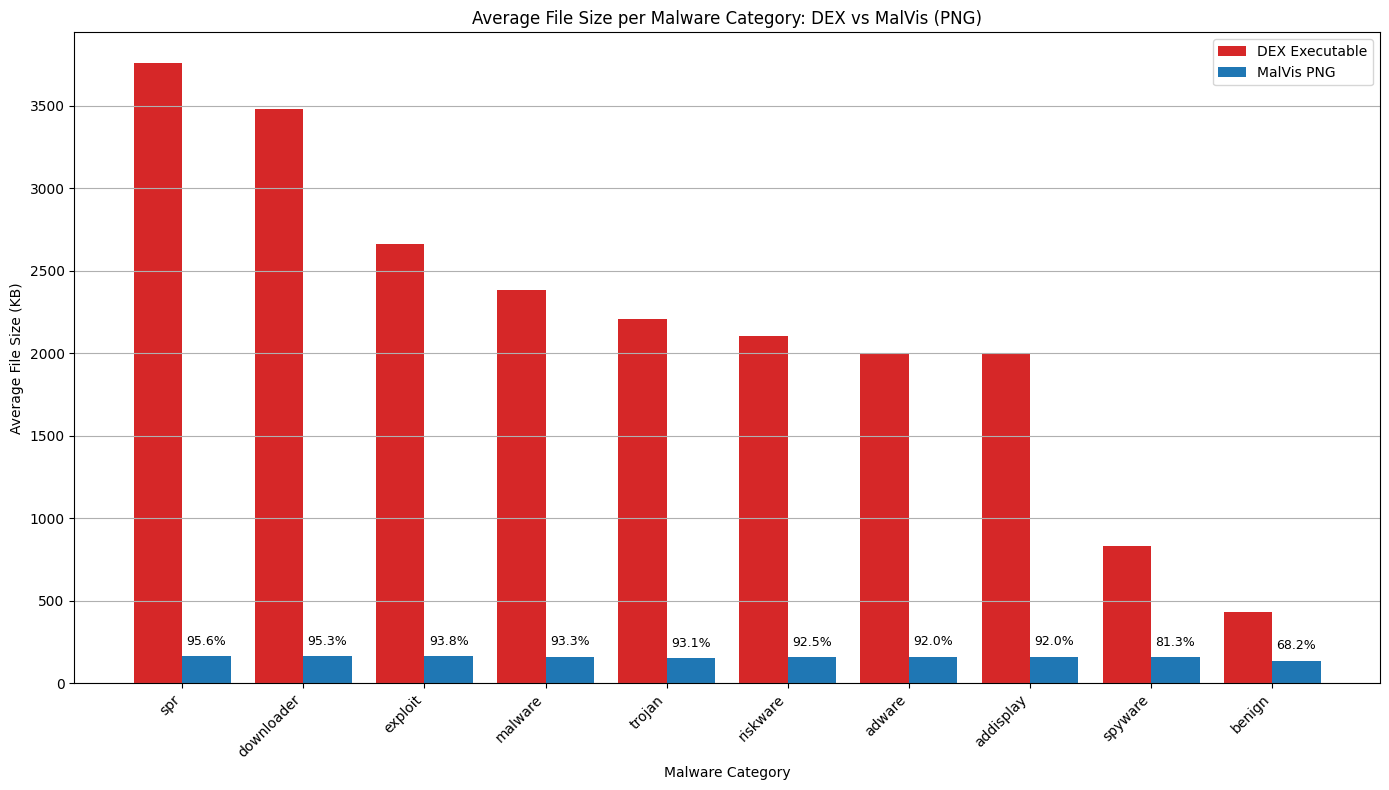}
  \caption{Comparison of average file sizes in DEX executables vs malVis PNG representations across malware categories (with size reduction percentages)}
  \label{fig:file sizes comparision}
\end{figure}

Deep Neural Networks, especially CNNs, have shown exceptional performance across domains such as vision, biomedical, and cybersecurity~\cite{dhillon2020convolutional,sun2025survey,alblwi2025d,sun2021anomaly}, primarily due to the availability of large, structured datasets. In malware detection, transforming code into image representations allows CNNs to identify visual patterns of malicious behavior, offering a scalable, non-executable, and efficient alternative to traditional analysis methods. These representations are significantly smaller than the raw executable files (Fig.~\ref{fig:file sizes comparision}), reducing storage needs and execution risks.

Our approach further benefits from transfer learning by leveraging pretrained CNNs trained on massive image datasets such as ImageNet, enabling effective pattern recognition in malware with minimal domain-specific training. Despite these advantages, progress is hindered by limited access to large-scale, interpretable, and public malware visualization datasets. Addressing this gap is essential for advancing robust, explainable, and reproducible malware detection research.

\subsection{Existing Malware Image Datasets}
We highlight two widely known Android malware datasets: AndroZoo~\cite{allix2016androzoo} and Drebin~\cite{arp2014drebin}, both commonly used in Android malware detection research. However, because our MalVis approach focuses on transforming bytecode into images for visualization, we primarily compare MalVis with other existing image-based datasets in this section.

Scott Freitas~\emph{et al.}~\cite{freitas2022malnet} introduced the MalNet database, a substantial contribution to the field with over 1.2 million malware images spanning 47 types and 696 families. While their direct byte-to-location color mapping method is innovative within the Android application structure, our dataset, MalVis, offers enhancements in the form of over 1.3 million images, with a particular focus on addressing malware obfuscation techniques. This focus improves the effectiveness of Android malware detection, as discussed in more detail by Makkawy~\emph{et al.}~\cite{makkawy2024improving}.

Virus-MNIST, proposed by David A.~\emph{et al.}~\cite{noever2021virus}, is a large publicly available malware image dataset. The dataset includes 51,880 grayscale images of malware, classified into nine virus classes and one benign class, all formatted as \([32\times32]\) images. The dataset represents the malware classification problem, like the famous MNIST dataset used for handwriting recognition. Malware images are generated by converting the first 1,024 bytes of Portable Executable (PE) files into \([32\times32]\) grayscale images. Although Virus-MNIST introduces a significant step towards standardizing malware image datasets, its representation of malware using only the first 1,024 bytes may result in a lack of capturing the complete characteristics of the malware~\cite{rezaei2021pe}.

L. Nataraj~\emph{et al.}~\cite{nataraj2011malware} presents the MalImg dataset, which offers a straightforward and efficient malware visualization and classification method. Using image processing techniques, this approach classifies malware samples based on their similarity to specific malware types, utilizing standard image features. MalImg achieves a notable classification accuracy of 98\% on a dataset comprising 9,458 samples across 25 distinct malware types. However, with this limited dataset size, there is a possibility that the model is overfitting to the specific characteristics of these samples. 

Table~\ref{tab:datasets} summarizes public and private image-based malware datasets, including MalVis, MALNET-IMAGE, and Virus-MNIST, providing details on the number of classes and dataset size. Despite existing visualization contributions, these methods face some limitations. As observed by Kunwar~\emph{et al.}~\cite{kunwar2024sok}, MalNet encodes malware bytecode based on byte location in the executable file, lacks resilience to obfuscation, and does not identify suspicious behaviors. Virus-MNIST~\cite{noever2021virus} uses only the first 1,024 bytes of PE files, limiting its representational scope. In contrast, MalVis combines entropy and N-gram analysis to generate color-encoded RGB representations that emphasize abnormal structures, encryption, packing, and compression behaviors. Our experiments show that this richer visual encoding enhances model interpretability and improves classification performance. A detailed discussion of the MalVis dataset and its visualization approach is presented in Section~\ref{sec:method}.

\begin{table}[ht]
\centering
\caption{Summary of image-based malware datasets detailing the number of classes,  dataset sizes, and availability.}
\label{tab:datasets}
\begin{tabular}{lccccc}
\hline
\textbf{Dataset}      & \textbf{\# Classes}  & \textbf{Dataset Size} & \textbf{Public} & \textbf{Private} \\ \hline 
\rowcolor{LightBlue}\textbf{MalVis}& 10          & 1,300,822            & \checkmark              &                  \\
\textbf{MalNet}~\cite{freitas2022malnet} & 696          & 1,262,024            & \checkmark              &                  \\
\textbf{AndroDex}~\cite{aurangzeb2024androdex} & 180          & 24,746            & \checkmark              &                  \\

\textbf{Virus-MNIST}~\cite{noever2021virus}  & 10              & 51,880               & \checkmark              &                  \\
\textbf{Malimg}~\cite{kalash2018malware}     & 25               & 9,458                & \checkmark              &                  \\ 
\cdashline{1-5}[1pt/1pt]
\textbf{Microsoft}~\cite{malware-classification} & 9            & 108,000              &                & \checkmark                \\ 
\textbf{IVMD-2013}~\cite{kancherla2013image}& 2            & 37,000              &                & \checkmark                \\ 
\textbf{AdvAndMal}~\cite{wang2021advandmal}
& 12            & 5,560              &                & \checkmark                \\ 
\textbf{Halil-2020}~\cite{unver2020android}
& 2            & 29,100            &                & \checkmark                \\ 
\hline
\end{tabular}
\end{table}

\subsection{Visualization Strategies for Malware Detection}
As image-based malware detection has become a powerful paradigm for analyzing Android applications, it bypasses manual feature engineering through automated visual pattern recognition. Current approaches primarily focus on two types of representations:

\subsubsection{Grayscale Image Encoding}
The foundational work by Nataraj~\emph{et al.}~\cite{nataraj2011malware} established grayscale conversion by mapping binary bytes to pixel values in the range (0–255), revealing structural patterns in malware families. Modern implementation includes DexRay by Nadia Daoudi~\emph{et al.}~\cite{daoudi2021dexray}; it converts DEX bytecode into 1D grayscale vectors (1×128×128) for CNN classification, achieving 96\% F1-score while resisting obfuscation. Despite its highest accuracy, their approach resulted in smaller-size grayscale images that could be affected by more data loss in the representations. In another instance, Wang~\emph{et al.}~\cite{wang2021novel} developed a novel scheme that combines static and dynamic analysis with CNN for efficient malware detection and classification. Their method integrates a Convolutional Block Attention Module (CBAM) with CNN to detect malware similarities using grayscale images from the MalImg and Microsoft datasets. However, their experiment was conducted on a relatively small dataset of approximately 20,000 samples covering 25 types of malware, which could be affected by model overfitting.

\subsubsection{RGB Image Encoding}
Advanced malware variants often exhibit more sophisticated patterns and behaviors that are difficult to capture using standard grayscale or single-channel representations. Additional color channels are required to encode these complex characteristics, enabling richer feature representation and enhancing the model's ability to detect subtle malicious traits. Asim~\emph{et al.}~\cite{darwaish2020rgb} introduced a technique that transforms APK files into lightweight RGB images utilizing a predefined dictionary and an intelligent mapping mechanism. Their method converts the \textit{AndroidManifest.xml} permissions into ASCII values, which are then aggregated and encoded into a single color value. 

While this approach facilitates the image-based representation of APK features in RGB channels, it suffers from significant information loss due to the summation of ASCII values, which flattens each permission into a single numerical value. This reduction hinders the model's ability to capture detailed permission information, thus limiting its effectiveness in capturing nuanced malicious behaviors.

Progress in this field is hindered by the scarcity of publicly available visualized malware datasets~\cite{ismail2024malssl} and the need for robust methodologies to capture malware patterns and behaviors~\cite{freitas2022malnet} effectively. The MalVis dataset aims to tackle these challenges by providing comprehensive representations of malware, which convert abnormal operational and structural patterns in bytecode into visual forms. In addition, it includes multiclass labels for accurate malware classification and analysis, thereby improving targeted threat identification.
\section{Methodology}\label{sec:method}
This section describes the methodology for collecting and constructing the MalVis dataset, including data generation and construction, MalVis bytecode-to-image visualization, CNN architectures with experiment settings, and environment setup.

\subsection{Data Generation and Construction}\label{sec:data-gen}
\begin{figure}[h]
  \centering
  \includegraphics[width=130mm]{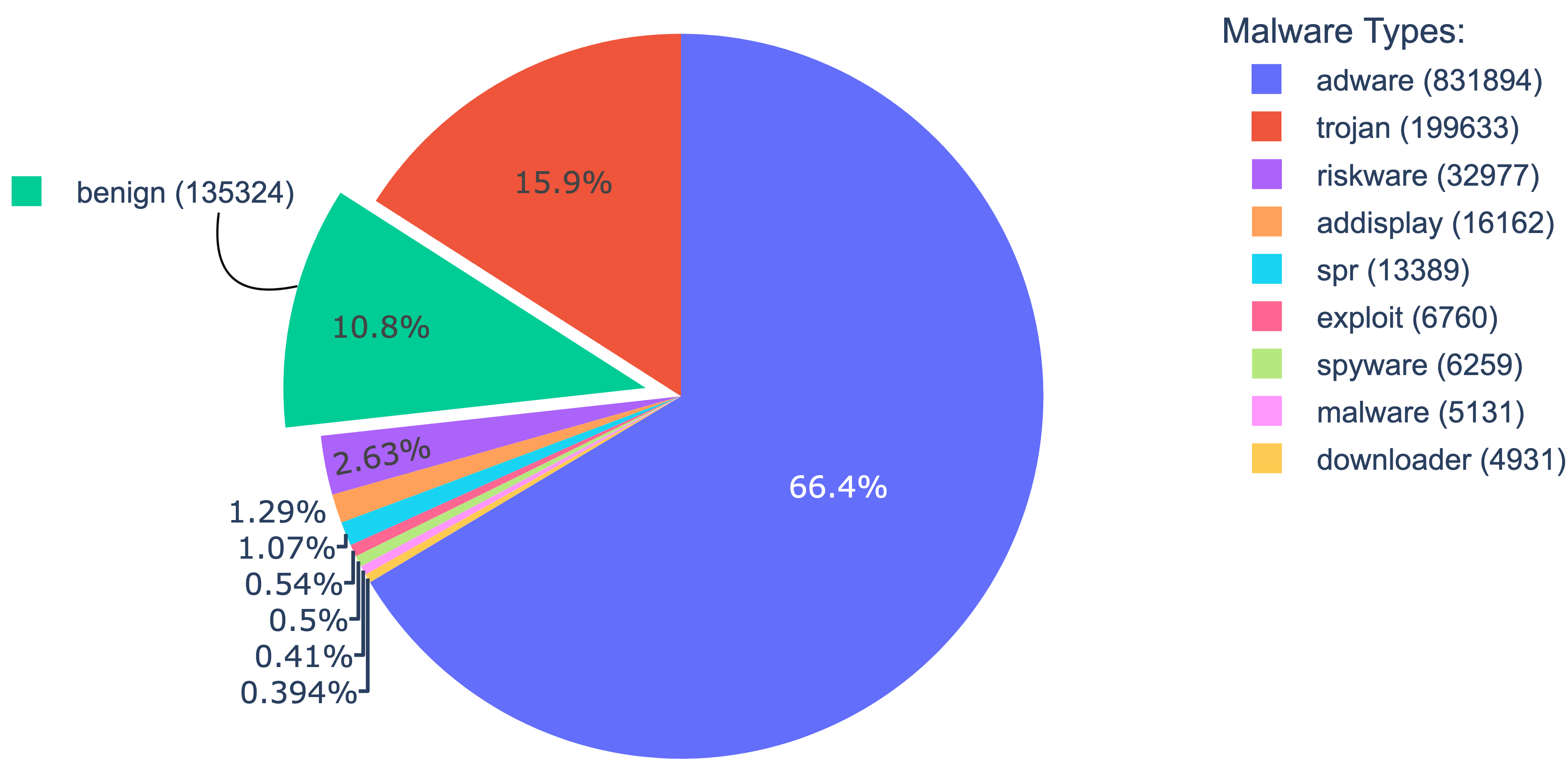}
  \caption{Distribution of malware types and benign in MalVis.}
  \label{fig:distribution_of_samples_in_MalVis}
\end{figure}

The MalVis generation process uses a subset of the AndroZoo dataset~\cite{allix2016androzoo}, a key resource in Android research encompassing 24,743,375 applications collected from platforms like the Google Play Store. The binary classification dataset consists of 49,150 malware and 135,324 benign samples. The multiclass malware dataset utilizes Euphony~\cite{hurier2017euphony} to categorize malware into 289 distinct classes. For training purposes, we focus on the nine largest classes to enhance labeling accuracy and reduce false positives, avoiding samples with multiple labels. We also verify these samples using Virustotal~\cite{virustotal} to ensure reliability. The refined dataset primarily included malware samples, along with an addition of 135,324 benign samples sourced from AndroZoo. Figure~\ref{fig:distribution_of_samples_in_MalVis} presents the distribution of nine malware and the benign class within 1,300,822 application visualizations.

\subsection{MalVis bytecode-to-image visualization}
\begin{figure}[ht]
    \centering
    \includegraphics[height=160mm,width=\linewidth]{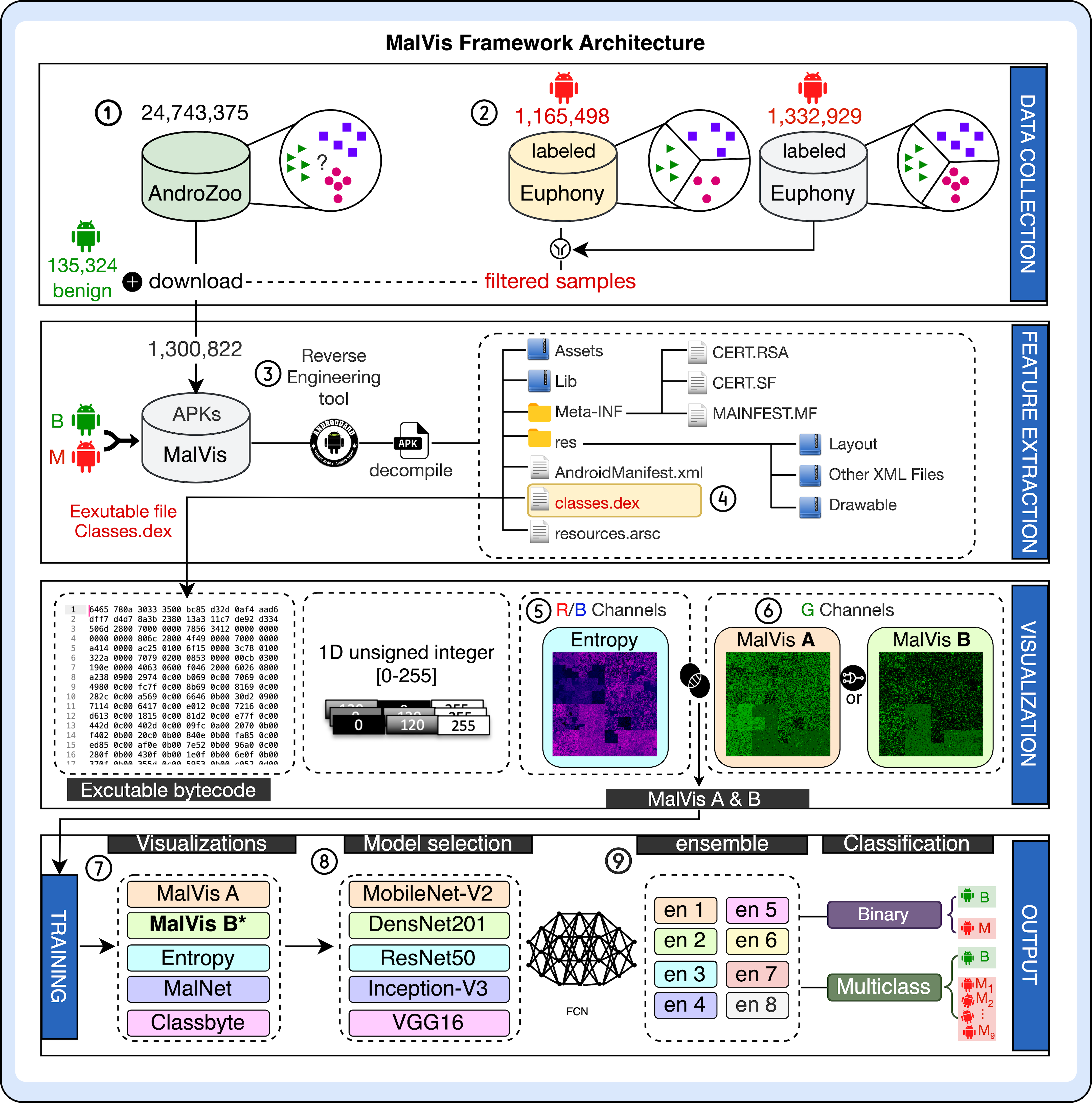}
    \caption{A schematic illustration of the proposed framework architecture is organized into four distinct rows. The first row details the data collection and labeling process. The second row focuses on feature extraction. The third row constructs and generates RGB images using entropy and N-gram methods. The bottom row describes the training process, including visualization techniques, CNN models, ensemble methods, and both binary and multiclass classifications.}
    \label{fig:structure}
\end{figure}

The MalVis bytecode-to-image visualization process begins by extracting Dalvik Executable (DEX) files from Android APKs using AndroGuard~\cite{AndroGuard}, a well-known reverse engineering tool. This step yields the \texttt{classes.dex} files, as illustrated in Fig.~\ref{fig:structure}\circled{4}. These \texttt{.dex} files consist of byte values in the range of \texttt{0x00} to \texttt{0xFF}. These values are first converted into a one-dimensional array of unsigned integers, where each value is between 0 and 255. These integers correspond directly to the pixel color intensities. The 1D array is reshaped into a two-dimensional grayscale image with a fixed width and height of 256 pixels to visualize the bytecode. This transformation employs Nearest-Neighbor Interpolation (NNI) and the Pillow library in Python, ensuring consistent image dimensions while preserving the original byte sequence structure.

Shannon entropy is applied to the executable dex file using a 32-byte sliding window to determine the red and blue channels. These channels are defined by distinct formulas motivated by ~\cite{cortesi_scurve}, as described in our earlier paper~\cite{makkawy2024improving}. Each formula utilizes Shannon entropy \ref{form: Shannon Entropy} differently to highlight regions of considerable randomness, which may indicate encryption or obfuscation. Note
\begin{equation}
\label{form: Shannon Entropy}
H(X) = -\sum_{i=1}^{N} P(x_i)\log_{2}(P(x_i)) ,
\end{equation}
where
\begin{itemize}
    \item {$H(X)$}: Represents the Shannon entropy of the random variable {$X$}, which measures the uncertainty or randomness in the 32-byte sequence.
    \item {$P(x_i)$}: The probability of observing the specific {$i^{th}$} outcome or byte value {$x_i$}. It is calculated as the frequency of {$x_i$} in the 32-byte sequence divided by the total number of bytes (32).
    \item {$N$}: Denotes the total number of unique outcomes for the random variable {$X$}. For a single byte, {$N = 256$} (corresponding to values {$\{0, 1, 2, ..., 255\}$}).
    \item {$x_i$}: Refers to a specific byte value in the range {$\{0, 1, 2, ..., 255\}$}. In the context of a sliding window of 32 bytes, it represents the {$i^{th}$} byte in the sequence.
    \item {$\log_{2}$}: The logarithm to base two is commonly used in entropy calculations to express the result in units of bits.
\end{itemize}
Given the varied types of malware introduced by the MalVis dataset, we have explored techniques to improve the recognition of these variations. Our analysis focuses on extending our earlier approach~\cite{makkawy2024improving} from two color channels (red and blue) to three channels by encoding additional feature into the green channel of RGB images using two primary encoding methods:

\begin{itemize}
    \item \textbf{Classbyte Encoding:} We adopt the Classbyte encoder introduced by Duc-Ly~\emph{et al.}~\cite{vu2020hit4mal}, which maps semantic features of bytecode to varying intensities of the green channel. We selected this method due to its effectiveness and comparable performance to our previously employed entropy-based encoding for binary classification tasks.
    
    \item \textbf{N-gram Structural Encoding:} We incorporate N-gram representations derived from byte sequences to capture the malware bytecode's underlying structural patterns and contextual dependencies. This technique, commonly used in malware detection research~\cite{ali2020malgra,zhong2024enhancing}, enriches the green channel with statistical features that reflect code regularities and anomalies, thereby enhancing the capability to distinguish between different malware types. 
\end{itemize}

The following subsections discuss the implications of these methods for advancing malware visualization.

\subsubsection{Approach MalVis-A}
\begin{figure}[h]
    \centering
    \includegraphics[width=130mm]{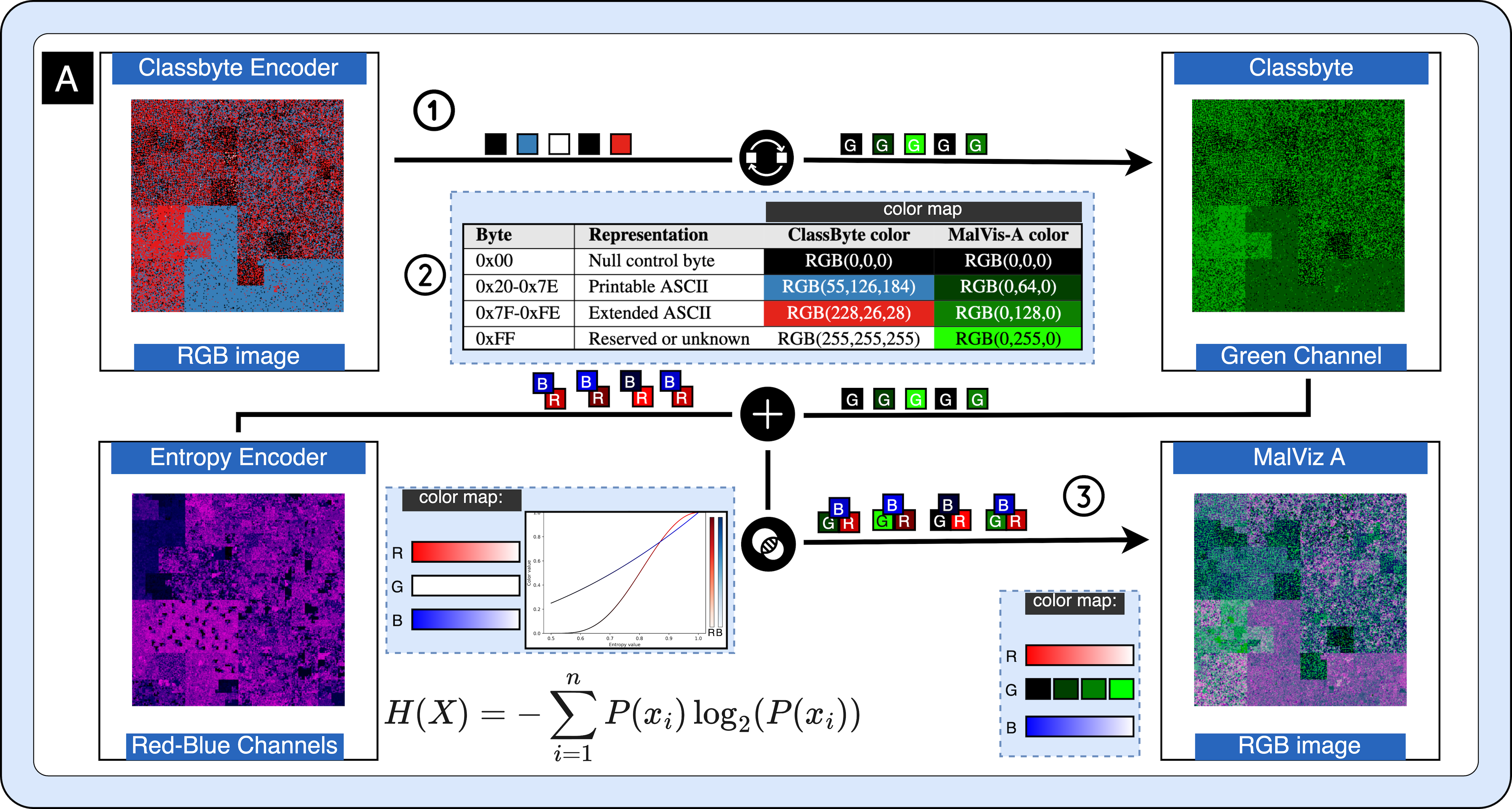}
    \caption{Overview of constructing the MalVis-A visualization method, resulting in RGB image representations using the Classbyte encoding in the green channel and encoding entropy in the red and blue channels.}
    \label{fig:structure A}
\end{figure}

This approach uses the Classbyte representation, which performs similarly to the entropy encoder in binary classification. It translates the features identified by the four Classbyte colors into four distinct shades of green in the green channel, as illustrated in Fig.~\ref{fig:structure A}~\circled{1}. The method highlights sections of bytecode containing both clear-text printable and non-printable ASCII characters and null byte areas, as illustrated in Fig.~\ref{fig:structure A}~\circled{2}. These distinctions assist in analyzing the bytecode to determine whether it has been encrypted or injected with null bytes to evade malware detection. 

The previously generated red and blue channels are combined with the newly constructed green channel, resulting in MalVis-A RGB images, as shown in Fig.~\ref{fig:structure A}~\circled{3}. Unfortunately, this approach did not yield the desired improvement in the accuracy of multiclass classification. Further results of the analysis and evaluation of this approach are presented in Section~\ref{sec:results}.

\subsubsection{Approach MalVis-B}
\begin{figure}[h]
    \centering
    \includegraphics[width=130mm]{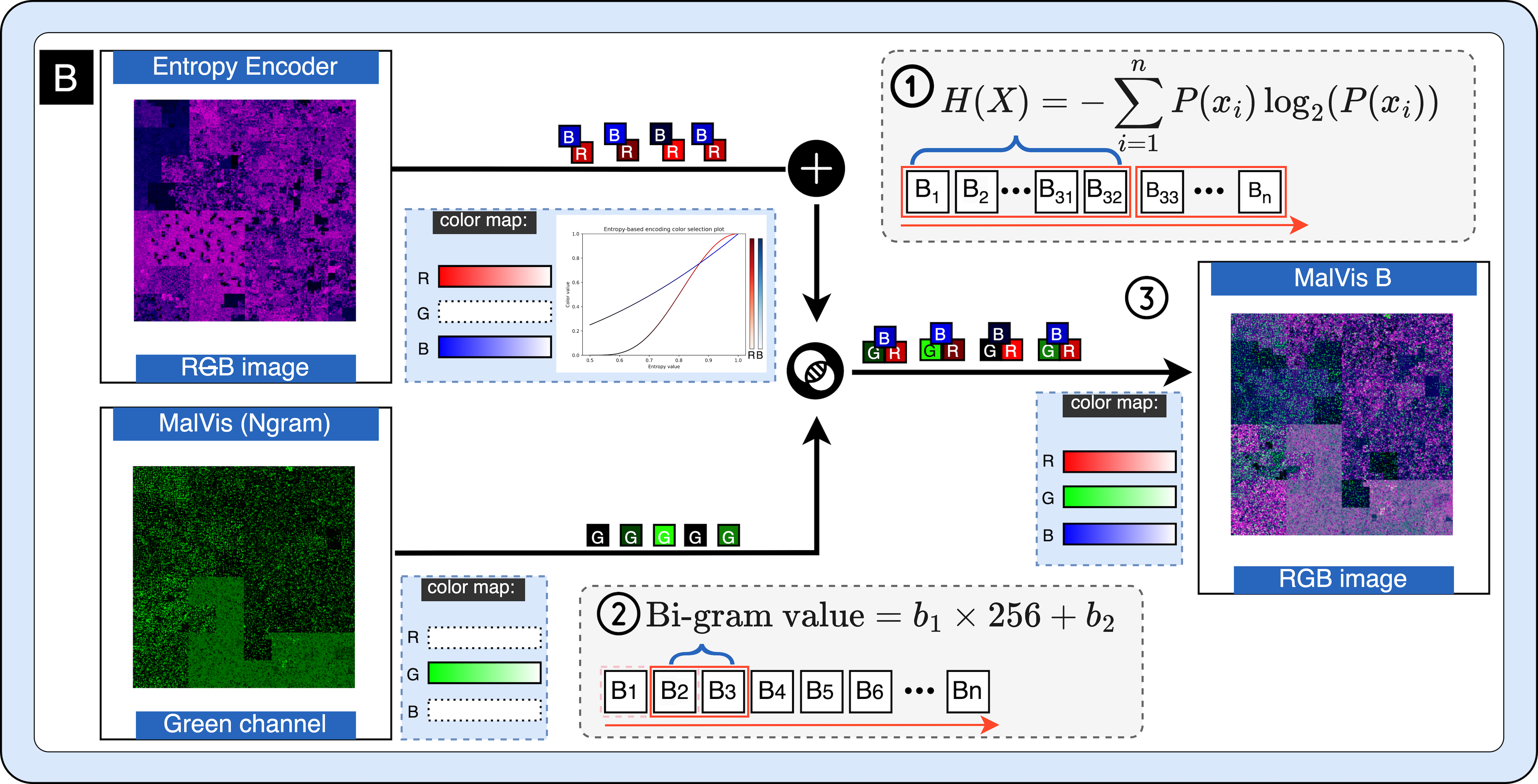}
    \caption{Overview of constructing the  MalVis-B visualization method using the N-gram encoding in the green RGB channel.}
    \label{fig:structure B}
\end{figure}

This approach utilizes the N-gram method, which has been extensively studied for malware anomaly detection. The approach is particularly relevant for Android applications, which are often written in Java and Kotlin, thus inheriting the programmatic structure. Abnormalities are detected when the byte sequences differ from the typical bytecode structure using the green channel depicted in Fig.~\ref{fig:structure B}~\circled{2}. One of the key goals of MalVis is to bridge the gap between raw visualization and interpretability. Unlike prior methods that map byte values to color values without semantic linkage, our framework encodes interpretable attributes: entropy highlights encrypted or compressed code regions. At the same time, N-gram transitions emphasize structural irregularities in bytecode. This mapping allows security analysts and researchers to visually associate distinct color patterns with specific malware behaviors, such as repacking functions or obfuscation. 

Figures (\ref{Java example}-\ref{Java Bi-gram}) further clarify this concept. For example,  the script in Fig.~\ref{Java example} below depicts a simple Java code for a \textit{'for'} loop before it is compiled into bytecode. The bytecode often reveals specific patterns that represent the underlying syntax and structure of the program. The keyword \textit{'for'}
indicates the presence of a loop followed by an initialization statement, a condition, and an increment surrounded by braces $(\cdots )$, while curly braces $\{ \cdots \}$ denote the loop's body. 

\begin{figure}[t]
    \begin{lstlisting}
    // For_Loop.java
    public class Main {
      public static void main(String[] args) {
        for (int i = 0; i < 5; i++) {
            System.out.println(i);
        }  
    }
    
    \end{lstlisting}
    \caption{An example of a simple for-loop written in Java.}
    \label{Java example}
\end{figure}

Running "javac For\_Loop.java" compiles the code into a bytecode file named "For\_Loop.class", which the DVM uses to execute the program, as demonstrated in Fig.~\ref{Java bytecode}.

\begin{figure}[t]
    \begin{lstlisting}
    // For_Loop.class
    iconst_0            // Push 0 into the stack.
    istore_1            // Store 0 to local variable 1.
    LoopStart:
    iload_1             // Load variable 1 (i) into the stack.
    iconst_5            // Push 5 into the stack.
    if_icmpge EndLoop   // Compare i and 5, jump to EndLoop if i >= 5.
    getstatic           // Access System.out.
    iload_1             // Load variable 1 (i) into the stack.
    invokevirtual       // Call System.out.println(i).
    iinc 1, 1           // Increment variable 1 (i++).
    goto LoopStart      // Jump to the start of the loop.
    EndLoop:
    \end{lstlisting}
    \caption{Translation of a Java for-loop into its equivalent JVM instructions in bytecode form after compilation.}
    \label{Java bytecode}
\end{figure}
As illustrated by the MalVis-B algorithm in Fig.~\ref{alg:MalVis algorithm}, we implemented the Bi-gram method on the raw bytes using a two-byte window to recognize anomalies in the bytecode's operational structure. Using Bi-gram on the bytecode represents the transition between instructions as listed in Fig.~\ref{Java Bi-gram}.
\begin{figure}[ht!]
    \begin{lstlisting}
    iconst_0 istore_1  // Initialization of i to 0.
    istore_1 iload_1 // Load the stored value of i.
    iconst_5 if_icmpge //Conditional jump.
    getstatic iloa_1 // Prepare to print the value of i.
    invokevirtual iinc // After printing, increment i.
    iinc goto // Jump back to the start of the loop.
    \end{lstlisting}
    \caption{Representation of the employed Bi-gram approach on the Java instructions capturing the semantic transition of these instructions.}
    \label{Java Bi-gram}
\end{figure}

The Bi-gram method can detect obfuscated code by identifying irregular Bi-gram patterns. A two-byte window size allows for detecting simple structural patterns, while larger windows capture more complex ones. However, this decision involves a trade-off that requires more computational resources and time to generate over 1.3 million images. Given limited resources, this study used a two-byte window size. Further research could optimize the window size for improved performance.

The Bi-gram formula is 
\begin{equation}\label{form: Bi-gram formula}
    \text{Bi-gram value} = b_1 \times  2^8 + b_2 ,
\end{equation}
which takes two consecutive bytes,~{$b_1$} and~{$b_2$}, to compute the Bi-gram value. The multiplication of the first byte~{$b_1$} by {$2^8=256$} shifts it to higher-order in the combined value, which is then added to the~{$b_2$} value, as shown in line 16 of the Algorithm, Figure~\ref{alg:MalVis algorithm}. The resulting Bi-gram value represents the degree level of a green pixel and is normalized to the range $[0, 1]$ by dividing by the maximum possible value of {$(256\times256)-1=65,535$} as described by
\begin{equation}\label{form: Bi-gram g formula}
    g = \frac{\text{Bi-gram value}}{(256\times256)-1} = \frac{b_1 \times 256 + b_2}{65,535}.
\end{equation}

Finally, if the byte is the last in the file, it is reset to 0, as detailed in lines 18 and 19 of the Algorithm, Fig.~\ref{alg:MalVis algorithm}. Hence, MalVis presents a conceptually innovative visualization design that maps meaningful malware properties to distinct visual domains. Consequently, MalVis is more effective and better aligned with the objectives of explainable malware classification. This approach has demonstrated improved accuracy in the context of multiclass MalVis, and both representation techniques are evaluated in Section~\ref{sec:results}.

\begin{figure}[H]
    \centering
    \begin{algorithm}[H]
        \footnotesize
        \caption{\textbf{MalVis-B} Visualization Algorithm}\label{alg:entropy_ngram}
        \begin{algorithmic}[1]
            \State \textbf{Input:} Data array $\mathit{data}$ of bytecode, symbol map $\mathit{symbol\_map}$, index $x$
            \State \textbf{Output:} RGB values in the range [0, 255]
            \State $e \gets \text{Entropy}(\mathit{data}, 32, x, \text{len}(\mathit{symbol\_map}))$
            \Comment\textcolor{dkgreen}{Calculate entropy using a window size of 32 bytes}
            \Function{curve}{$v$}
                \State $f \gets (4v - 4v^2)^4$
                \State $f \gets \max(f, 0)$
                \State \Return $f$
            \EndFunction
            \If{$e > 0.5$}
                \State $r \gets \text{curve}(e - 0.5)$
                \Comment\textcolor{dkgreen}{Red component is determined by the scaled entropy value}
            \Else
                \State $r \gets 0$
                \Comment\textcolor{dkgreen}{If entropy is less than or equal to 0.5, set red component to 0}
            \EndIf
            \State $b \gets e^2$
            \Comment\textcolor{dkgreen}{Blue component is proportional to the square of entropy}
            \If{$x < \text{len}(\mathit{data}) - 1$}
                \State $n\_gram\_value \gets (\mathit{data}[x] \ll 8) + \mathit{data}[x + 1]$
                \Comment\textcolor{dkgreen}{Compute 2-byte n-gram value}
                \State $g \gets \frac{n\_gram\ value}{65535}$
                \Comment\textcolor{dkgreen}{Normalize n-gram value to [0, 1] for green component}
            \Else
                \State $g \gets 0$
                \Comment\textcolor{dkgreen}{If at the last byte, the green component is set to 0}
            \EndIf
            \State \Return $[ \text{int}(255 \cdot r), \text{int}(255 \cdot g), \text{int}(255 \cdot b) ]$
            \Comment\textcolor{dkgreen}{Return RGB values scaled to the range [0, 255]}
        \end{algorithmic}
    \end{algorithm}
     \caption{Algorithm illustrating the generation of RGB image representation in the MalVis-B approach, utilizing entropy for the red and blue channels and N-gram for the green channel.}
    \label{alg:MalVis algorithm}
\end{figure}

\subsection{The Impact of Entropy and N-gram on MalVis representations experiments}
In this section, we further investigate the sensitivity and interpretability of MalVis visualizations. We conducted a controlled analysis by applying targeted transformations to a benign Android application, specifically WhatsApp \textit{Classes.dex}. The goal was to investigate and quantify the impact of encryption and unstructured operations in bytecode changes on the RGB image representations generated by our framework.

\subsubsection{Obfuscation Detection captured by Entropy in Red and Blue Channels}
\begin{figure}[H]
  \centering
  \includegraphics[width=130mm]{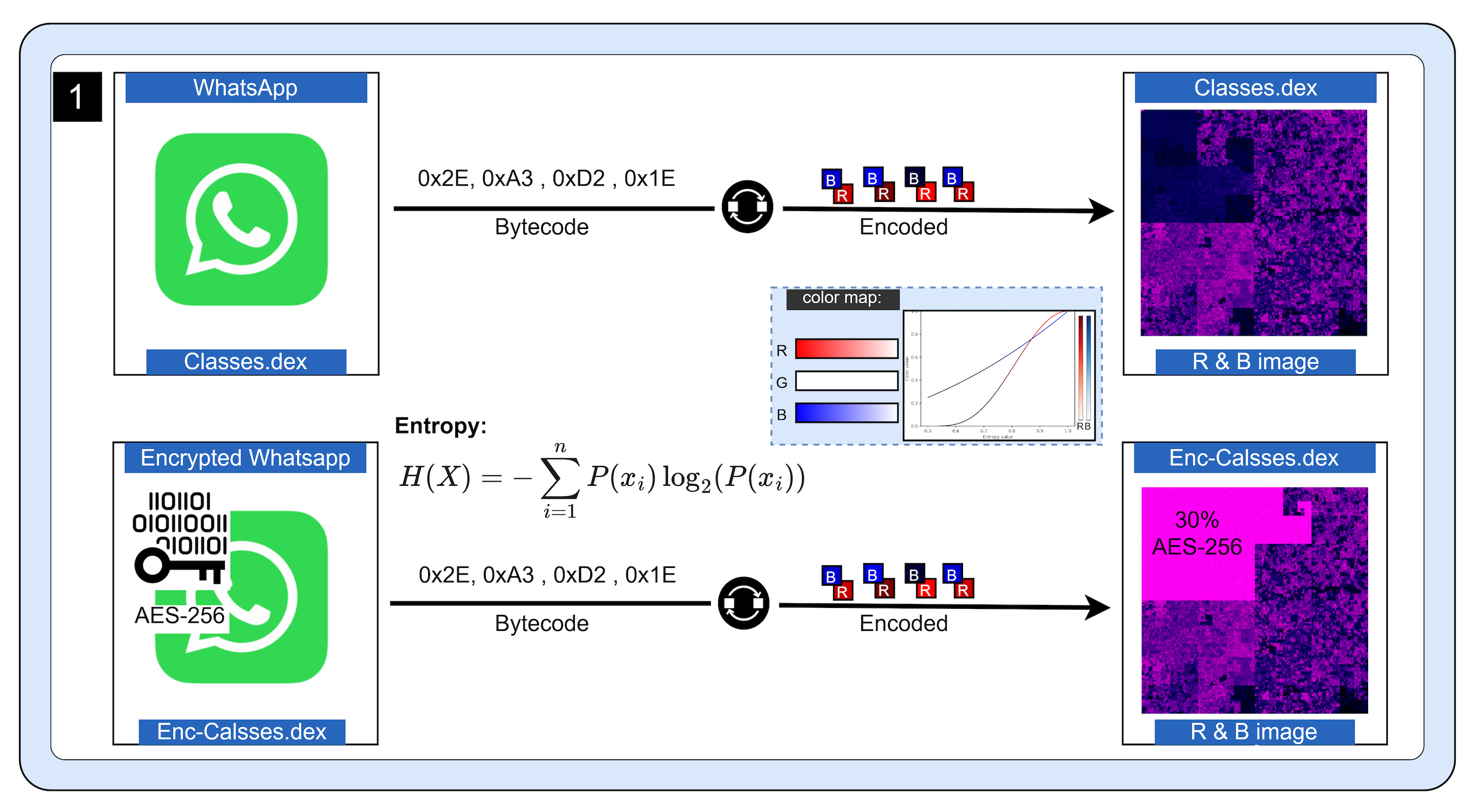}
  \caption{The impact of 30\% AES-256 encryption on \textit{Classes.dex} file captured by the entropy encoder in the red and blue channels of MalVis representations.}
  \label{fig: entropy on the red and blue channels.}
\end{figure}

In this experiment, we applied \texttt{AES-256} encryption in Electronic Codebook (ECB) mode to the initial 30\% of the \texttt{Classes.dex} bytecode. This encryption caused a noticeable entropy shift, particularly affecting image representations' red and blue channels. Entropy, which quantifies randomness over 32-byte windows, increased significantly in high-entropy areas, leading to brighter pixel intensities. This effect simulates obfuscation techniques that malware creators use to evade detection. As a result, the red and blue channels in Fig.~\ref{fig: entropy on the red and blue channels.} display brighter pixels in the top-left region, highlighting the encrypted sections in the representation.

\subsubsection{Unstructured bytecode Insertion captured by N-gram in Green Channel}
\begin{figure}[h]
  \centering
  \includegraphics[width=130mm]{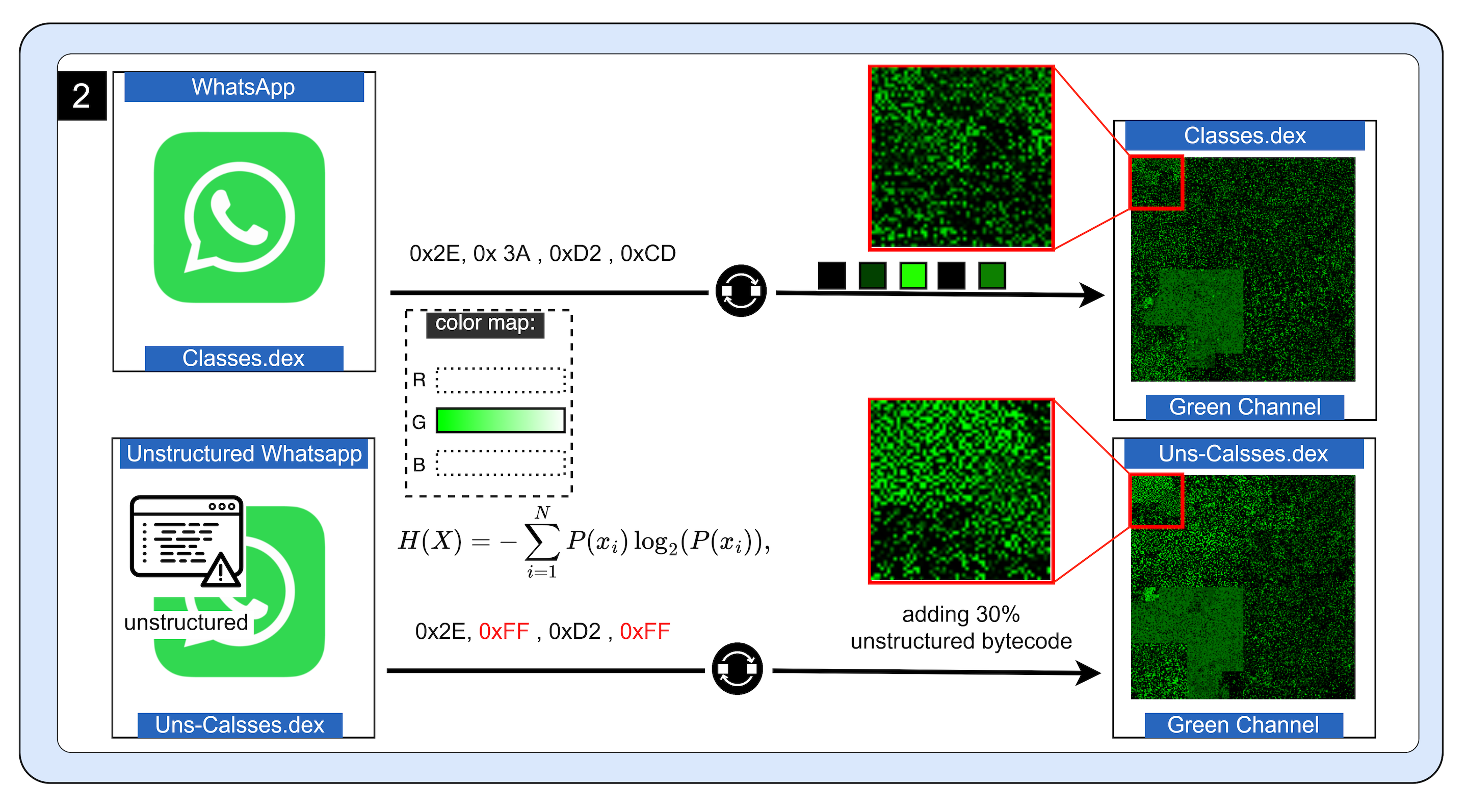}
  \caption{The impact of injecting 30\% randomized unstructured operations to \textit{Classes.dex} file captured by the N-gram encoder in the green channels of MalVis-B representations.}
  \label{fig: N-gram on the green channel.}
\end{figure}

In this experiment, we examined the structural sensitivity of the green channel by injecting random, unstructured operations into the initial 30\% of the \texttt{Classes.dex} file. This action disrupted the byte sequence, causing noticeable distortions in the N-gram values, significantly impacting the green channel. MalVis-B, which utilizes bi-gram formulas to detect abnormal operational patterns, recorded these disturbances as increased bi-gram values, resulting in brighter pixel values within green-channel textures, as depicted in Fig.~\ref{fig: N-gram on the green channel.}. These deviations were apparent when visualized next to an unchanged sample, highlighting the effectiveness of the green channel in detecting structural anomalies.

\subsection{Model Architecture and Experiment Settings}\label{sec:CNN architecture and experiment settings}
To assess the effectiveness of our proposed approach, we used a selection of well-recognized CNN models, including MobileNetV2, ResNet50, DenseNet201, VGG16, and InceptionV3. These models were applied to our generated visualizations and baseline comparison methods. The employed CNN models have proven highly effective in malware detection because they capture intricate patterns and features within image data~\cite{almomani2022automated, freitas2022malnet}. To ensure consistency across the CNN models, all images were resized to $224\times224$ pixels using nearest neighbor interpolation to align with the input dimensions required by the models. The dataset was partitioned into 80\% for training, 10\% for validation, and 10\% for testing. The batch size 64 was chosen based on empirical experimentation, as it provided an optimal trade-off between training speed and memory consumption on our GPU setup. The training was conducted over 50 epochs and carefully monitored to mitigate overfitting. This setup allowed for consistent and accurate assessments of the models' performance across different visualization techniques.

\subsection{Environment Setup}
MalVis visualization and model training were generated using an Ubuntu Server 22.04 LTS OS with x86 64 architecture. The hardware setup included a 16-core AMD Ryzen Threadripper PRO 5955WX processor, 128 GB of DDR4 RAM at 3200MHz, and an NVIDIA RTX A6000 graphics card. The system was configured within a controlled environment to ensure accurate results and minimize external influences.

\begin{table*}[t]
    \centering
        \caption{Comparison of visualization approaches using different CNN models. Abbreviations in the table include MNv2 (MobileNet-V2), DN201 (DenseNet201), RN50 (ResNet50), and INC-V3 (Inception-V3). Bold values highlight the highest score for each metric within the respective model. }
        \label{results-binary}
    \begin{tabular}{clccccccc}\toprule
        \textbf{ Approaches}                            & \textbf{ Models}  & \textbf{Accuracy} &\textbf{ F1-score} &\textbf{ Precsion} & \textbf{ Recall} & \textbf{ MCC} & \textbf{ R-AUC} \\\midrule
        \multirow{5}{*}{Classsbyte Encoder~\cite{vu2020hit4mal}}   & \textbf{MNv2}                  & 91.60\%                   & 85.42\%                    & 79.43\%                   & 92.39\%                   & 80.02\%                     & 96.91\%                       \\
                                                                      & \textbf{DN201}                 & 94.38\%                   & 89.39\%                    & 89.93\%                   & 88.85\%                   & 85.57\%                     & 97.57\%                       \\
                                                                      & \textbf{RN50}                  & 93.02\%                   & 86.51\%                    & 89.22\%                   & 83.97\%                   & 81.88\%                     & 96.38\%                       \\
                                                                      & \textbf{INC-V3}                & 94.38\%                   & 89.03\%                    & 92.73\%                   & 85.62\%                   & 85.38\%                     & 97.75\%                       \\
                                                                      & \textbf{VGG16}                 & 93.18\%                   & 87.33\%                    & 86.46\%                   & 88.22\%                   & 82.68\%                     & 96.17\%                       \\
                                                                     
\addlinespace[1.7ex]
\midrule
\addlinespace[1.7ex]
        \multirow{5}{*}{MalNet Encoder~\cite{freitas2022malnet}}       & \textbf{MNv2}                  & 92.78\%                   & 85.63\%                    & 91.16\%                   & 80.74\%                   & 81.10\%                     & 97.11\%                       \\
                                                                      & \textbf{DN201}                 & 89.66\%                   & 83.16\%                    & 74.46\%                   & 90.11\%                   & 77.27\%                     & 96.91\%                       \\
                                                                      & \textbf{RN50}                  & 86.34\%                   & 67.60\%                    & 91.92\%                   & 53.46\%                   & 63.22\%                     & 94.69\%                       \\
                                                                      & \textbf{INC-V3}                & 94.82\%                   & 90.01\%                    & 92.69\%                   & 87.47\%                   & 86.58\%                     & 97.84\%                       \\
                                                    & \textbf{VGG16}                 & 93.87\%                   & 88.02\%                    & 91.80\%                   & 84.54\%                   & 84.04\%                     & 97.38\%                       \\

\addlinespace[1.7ex]
\midrule
\addlinespace[1.7ex]
        \multirow{5}{*}{Entropy-based~\cite{makkawy2024improving}} & \textbf{MNv2}                  & 93.85\%                   & 88.10\%                    & 90.88\%                   & 85.50\%                   & 84.03\%                     & 97.57\%                       \\
                                                                      & \textbf{DN201}                 & \cellcolor{LightBlue}\textbf{95.32\%}                   & \cellcolor{LightBlue}\textbf{91.30\%}                    & 90.54\%                   & \cellcolor{LightBlue}\textbf{92.07\%}                   & \cellcolor{LightBlue}\textbf{88.10\% }                    & \cellcolor{LightBlue}\textbf{98.25\%}                       \\
                                                                      & \textbf{RN50}                  & 93.14\%                   & 86.50\%                    & 90.96\%                   & 82.47\%                   & 82.10\%                     & 97.11\%                       \\
                                                                      & \textbf{INC-V3}                & 94.94\%                   & 90.43\%                    & 91.12\%                   & 89.75\%                   & 86.99\%                     & 97.93\%                       \\
                                                                      & \textbf{VGG16}                 & 93.59\%                   & 87.00\%                    & \cellcolor{LightBlue}\textbf{94.64\%}                   & 80.49\%                   & 83.25\%                     & 97.45\%                       \\
                                                                      
\addlinespace[1.7ex]
\midrule
\addlinespace[1.7ex]
       \multirow{5}{*}{MalVis-A}             & \textbf{MNv2}                  & 91.97\%                   & 84.05\%                    & 89.22\%                  & 79.45\%                   & 78.94\%                     & 96.76\%                       \\
                                            & \textbf{DN201}                 & 94.95\%                   & 90.28\%                    & 92.60\%                   & 88.08\%                   & 86.92\%                     & 98.15\%                       \\
                                            & \textbf{RN50}                  & 93.64\%                   & 87.47\%                    & 92.03\%                   & 83.34\%                   & 83.40\%                     & 97.38\%                       \\
                                            & \textbf{INC-V3}                & 94.18\%                   & 88.63\%                    & 92.50\%                   & 85.07\%                   & 84.87\%                     & 97.88\%                       \\
                                            & \textbf{VGG16}                 & 92.81\%                   & 86.38\%                    & 87.26\%                   & 85.52\%                   & 81.50\%                     & 96.89\%                       \\

\addlinespace[1.7ex]
\midrule
\addlinespace[1.7ex]
        \multirow{5}{*}{\textbf{MalVis-B$^*$}}           & \textbf{MNv2}                  & \cellcolor{LightBlue}\textbf{95.04\%}                   & \cellcolor{LightBlue}\textbf{90.57\% }                   & \cellcolor{LightBlue}\textbf{91.76\%}                   & \cellcolor{LightBlue}\textbf{89.42\%}                   & \cellcolor{LightBlue}\textbf{87.22\%}                     & \cellcolor{LightBlue}\textbf{98.14\%}                       \\
                                                    & \textbf{DN201}                 & 95.22\%                   & 90.91\%                    & \cellcolor{LightBlue}\textbf{92.16\%}                   & 89.69\%                   & 87.68\%                     & 98.19\%                       \\
                                                    & \textbf{RN50}                  & \cellcolor{LightBlue}\textbf{95.08\%}                   & \cellcolor{LightBlue}\textbf{90.60\%}                    & \cellcolor{LightBlue}\textbf{92.35\%}                   & \cellcolor{LightBlue}\textbf{88.91\%}                   & \cellcolor{LightBlue}\textbf{87.30\%}                     & \cellcolor{LightBlue}\textbf{98.18\%}                       \\
                                                    & \textbf{INC-V3}                & \cellcolor{LightBlue}\textbf{95.19\%}                   & \cellcolor{LightBlue}\textbf{90.81\%}                    & \cellcolor{LightBlue}\textbf{92.58\%}                   & \cellcolor{LightBlue}\textbf{89.10\%}                   & \cellcolor{LightBlue}\textbf{87.58\%}                     & \cellcolor{LightBlue}\textbf{98.06\%}                       \\

                                                                                                                          & \textbf{VGG16}                 & \cellcolor{LightBlue}\textbf{94.60\% }                  & \cellcolor{LightBlue}\textbf{89.89\% }                   & 89.68\%                   & \cellcolor{LightBlue}\textbf{90.11\%}                   & \cellcolor{LightBlue}\textbf{86.21\%}                     & \cellcolor{LightBlue}\textbf{97.90\% }                      \\

        \bottomrule
    \end{tabular}
    \tiny{\scriptsize{*}\text{The results denote the optimal proposed method, referred to as MalVis-B.}}
\end{table*}
\section{Performance Measures}\label{sec:evaluation}
To ensure fairness when comparing the visualization methods and evaluating our proposed approaches alongside the baseline methods presented in Table~\ref{results-binary}, we employed accuracy, precision, recall, ROC-AUC, and MCC as validation metrics in the binary classification context. Similarly, the same metrics were employed for consistent evaluation in multiclass classification, as demonstrated in Table~\ref{tab:results-multi}. The accuracy~(\ref{eq:ACC}) indicates the percentage of instances correctly identified among the entire set of samples. The F1-score~(\ref{eq:F1}) provides a harmonic mean of the model's precision and recall, accounting for false positives and false negatives. Precision~(\ref{eq:P}) refers to the proportion of true positives in relation to all positive predictions made. Recall~(\ref{eq:R}) denotes the fraction of actual positives correctly identified by the model. ROC-AUC measures the area under the receiver operating characteristic curve, highlighting the balance between sensitivity and specificity. The MCC~(\ref{eq:MCC}) serves as a metric to assess classification performance, factoring in true and false positives and true and false negatives. 

Accordingly,
\begin{align}
    \label{eq:ACC} \text{Accuracy} &= \frac{TP + TN}{TP + TN + FP + FN}, \\
    \label{eq:F1} \text{F1-score} &= 2\times \frac{P \times R}{P + R}, \\
    \label{eq:P} \text{Precision} &= \frac{TP}{TP+FP}, \\
    \label{eq:R} \text{Recall} &= \frac{TP}{TP+FN}, 
\end{align}
and
\begin{equation}
\label{eq:MCC}\text{MCC} = \frac{TP\times TN - FP\times FN}{\sqrt{(TP+FP)(TP+FN)(TN+FP)(TN+FN)}},
\end{equation}
where TP, TN, FP, and FN represent true positive, true negative, false positive, and false negative, respectively. 

\section{Results}\label{sec:results}
This section provides a comparative analysis of the performance of the newly introduced visualizations, MalVis-A and MalVis-B, compared to baseline methods, namely MalNet~\cite{freitas2022malnet} and Classbyte, as detailed in the following subsections:

\subsection{Evaluation of MalVis-A and MalVis-B Performance Compared to Other Methods on a Binary Classification Dataset}
\begin{figure}[ht]
    \centering
    \includegraphics[width=130mm]{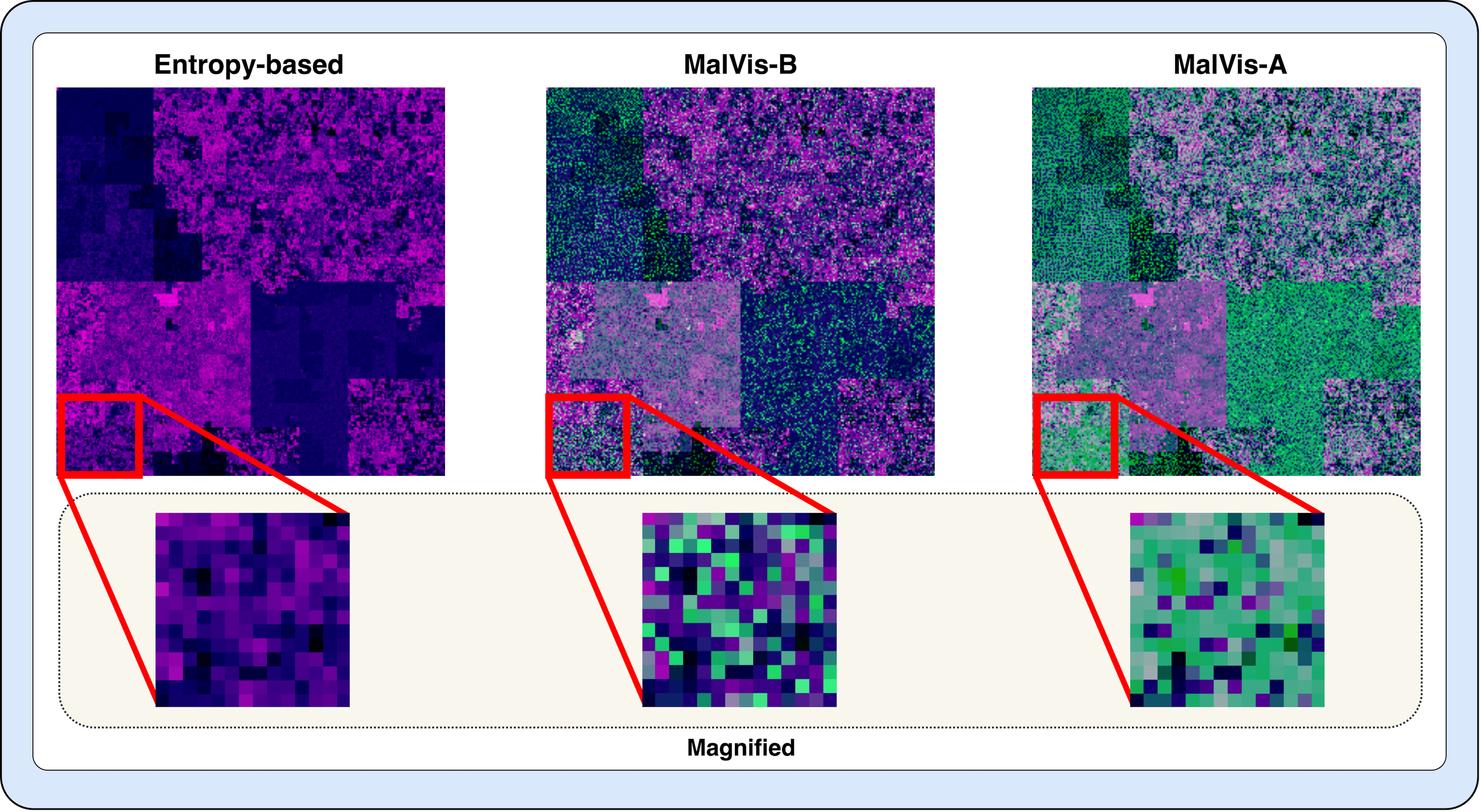}
    \caption{Figure showing the disruption caused by MalVis-A encoding of classbyte in the green channel, which impacts patterns captured by the entropy encoder compared to the representations by MalVis-B.}\label{fig:MalVis-A disruption}
\end{figure}

All methods used the same settings and were trained on identical subsets of training data to ensure a fair comparison. As shown in Table~\ref{results-binary}, the MalVis-A approach, which combines Classbyte and Entropy, did not improve classification performance as expected. Instead, it disrupted the patterns captured by the entropy encoder, as shown in Fig.~\ref{fig:MalVis-A disruption}. Encoding the four colors of the Classbyte method into a single green channel erases and replaces previously detected patterns. In contrast, the proposed MalVis-B method, which utilizes entropy and N-gram visualization, outperformed other methods in most CNN models except for DenseNet201. 

Although DenseNet201 did not show significant improvements across all metrics, it demonstrated superior precision, highlighting the model's effectiveness in accurately distinguishing true positives from false positives. The observed limitations in the remaining metrics are attributed to the highly imbalanced dataset outlined in Section~\ref{sec:data-gen}, where the prevalence of benign samples significantly exceeds that of malware. This imbalanced dataset was further addressed in the multiclass dataset evaluation detailed in Section~\ref{sec:undersampling malticlass}.


These experiments demonstrate that existing methods, including \textit{Classbyte} and \textit{MalNet}, provide limited semantic and structural variation, resulting in suboptimal performance for malware classification tasks. In contrast, \textit{MalVis-B} outperforms these approaches by integrating both entropy and N-gram patterns, producing meaningful visual representations that more effectively expose obfuscation, encryption, and other malicious behaviors. Notably, our earlier method relying solely on entropy~\cite{makkawy2024improving} did not achieve comparable performance, underscoring the value of combining multiple feature types. This highlights the need for enhanced visualization techniques that improve both interpretability and classification accuracy. Accordingly, \textit{MalVis-B} was selected for the subsequent advanced multiclass classification experiments to better distinguish between diverse malware types.

\subsection{Evaluation of MalVis-B Performance on Imbalanced Multiclass Dataset}
\begin{figure*}[p]
  \centering
  \includegraphics[height=0.93\textheight]{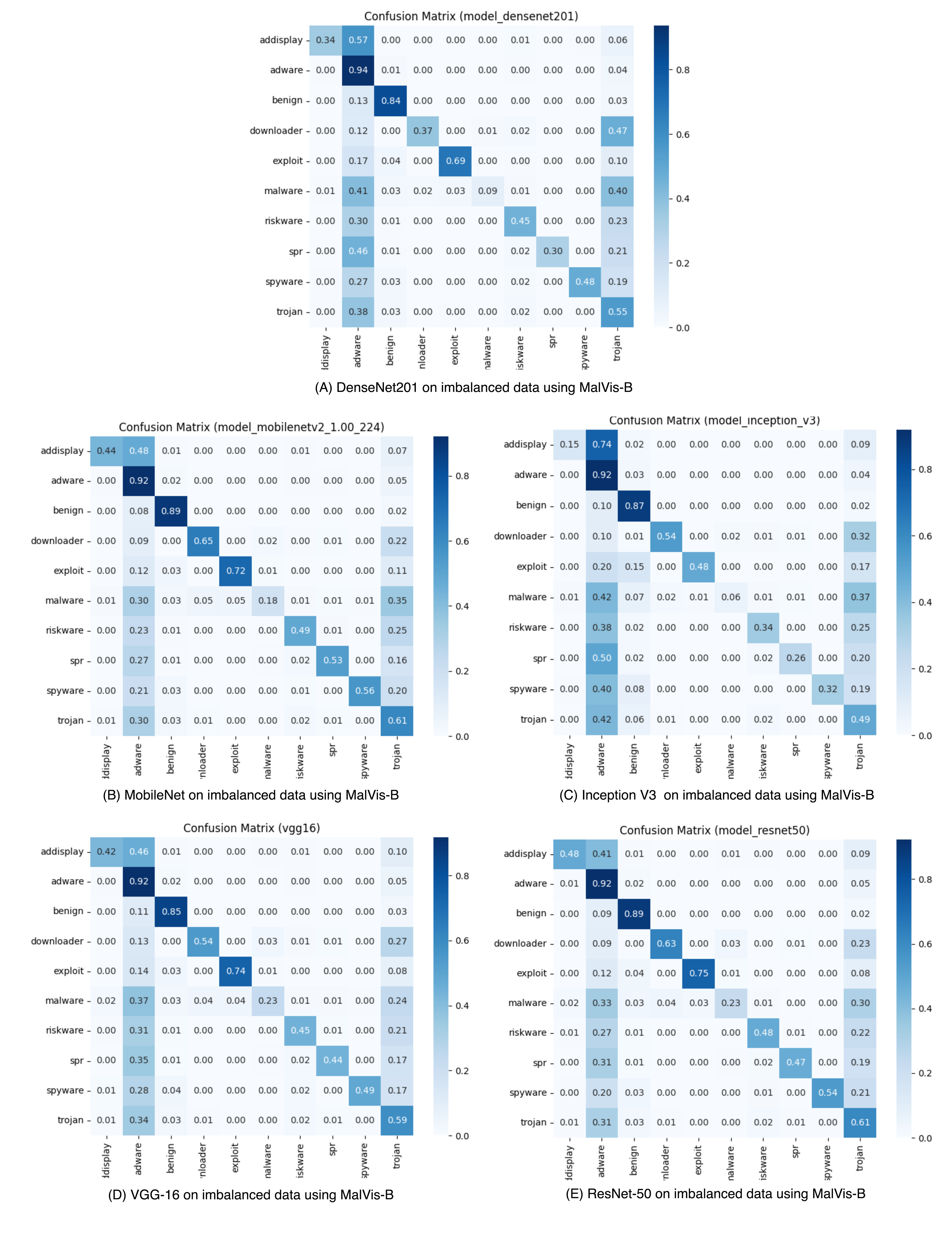}
  \caption{Confusion matrices of CNN models trained on the imbalanced multiclass MalVis dataset with the MalVis-B approach.}
  \label{fig:confusion imbalanced}
\end{figure*}

The evaluation of the MalVis-B representation in the imbalanced multiclass malware classification task, presented in Table~\ref{tab:results-multi}, demonstrated that the ResNet50 model achieved the highest performance. It achieved an overall accuracy of 94.03\%, F1-score of 83.54\%, and Precision of 83.34\%, surpassing the performance of state-of-the-art multiclass malware classification approaches~\cite{freitas2022malnet}. The analysis of the confusion matrix, presented in Fig.~\ref{fig:confusion imbalanced},~$\boxed{A}$ to $\boxed{E}$ reveals significant challenges in differentiating between the majority and minority classes within the imbalanced multiclass dataset. The darker column for the adware class suggests a bias due to its higher frequency in the training set, as shown in Fig.~\ref{fig:distribution_of_samples_in_MalVis}.

A deeper inspection of the confusion matrix Fig.~\ref{fig:confusion imbalanced} reveals frequent misclassification between Adware, Trojan, and Spyware classes. These malware types often share similar bytecode structures and use comparable obfuscation techniques, leading to visually overlapping patterns in the entropy and N-gram channels. For example, packed adware and spyware samples may exhibit high-entropy values with irregular n-gram sequences, which confound the classifier. These findings highlight the need for refined feature selection and possibly more semantic augmentation in future visualization efforts.
This highlights the effect of class imbalance, leading to biased decision boundaries that favor the majority class at the expense of consistent performance across all classes. 

Various strategies can address this imbalance, such as oversampling minority classes, undersampling majority classes, applying class weighting in the loss function, and using ensemble methods~\cite{mohammed2020machine,gosain2017handling}. The following sections cover applying undersampling to majority classes and discuss the evaluation of eight different ensemble methods in detail.

\begin{table}[ht]\centering
    \caption{Performance results of different models on the MalVis-B imbalanced multiclass dataset.}\label{tab:results-multi}
    \begin{tabular}{@{}lcccccc@{}}
        \toprule
        \multirow{2}{*}{\textbf{Models}} & \multicolumn{6}{c}{\textbf{MalVis-B Multi-Class}} \\
        \cmidrule(lr){2-7}
                                         & \textbf{A}            & \textbf{F1}         & \textbf{P}          & \textbf{R}          & \textbf{MCC}        & \textbf{ROC-AUC}     \\
        \midrule
        \textbf{MNv2}                    & 83.27\%               & 83.05\%             & 82.89\%             & 83.27\%             & 67.49\%             & 93.67\%              \\
        \textbf{DN201}                   & 82.81\%               & 81.81\%             & 81.74\%             & 82.81\%             & 65.27\%             & \cellcolor{LightBlue}\textbf{95.29\% }              \\
        \textbf{RN50}                    & \cellcolor{LightBlue}\textbf{84.03\% }              & \cellcolor{LightBlue}\textbf{83.54\% }             & \cellcolor{LightBlue}\textbf{83.34\% }            & \cellcolor{LightBlue}\textbf{84.03\% }            & \cellcolor{LightBlue}\textbf{68.53\% }             & 94.03\%              \\
        \textbf{INC-V3}                  & 80.18\%               & 78.67\%             & 78.74\%             & 80.18\%             & 59.95\%             & 93.39\%              \\
        \textbf{VGG16}                   & 82.56\%               & 81.99\%             & 81.72\%             & 82.56\%             & 65.42\%             & 91.87\%              \\
        \bottomrule
    \end{tabular}
\end{table}

\subsection{Evaluation of MalVis-B Performance using Undersampling on a Balanced Multiclass Dataset}\label{sec:undersampling malticlass}
We applied undersampling to the majority classes to address the imbalanced class distribution, creating a more balanced dataset. Although oversampling minority classes is the most effective data balancing method~\cite{mohammed2020machine}, we opted for undersampling due to limited computational resources and the time constraints associated with training the oversampled method. Table~\ref{tab:MalVis-B using undersampling on multi-class} presents the evaluation results for undersampling with MalVis-B. The confusion matrix in Fig.~\ref {fig:confusion matrix undersampleing}, models $\boxed{B}$ to $\boxed{F}$, highlights improved differentiation between majority and minority classes. Despite a 15-20\% reduction in overall performance relative to the results of the imbalanced dataset (Table~\ref{tab:results-multi}), we discuss ensemble methods to boost model performance in the following section.

\begin{table}[!ht]
\centering
    \caption{Performance results after undersampling approach on the large-MalVis dataset}\label{tab:MalVis-B using undersampling on multi-class}
    \begin{tabular}{@{}lcccccc@{}}
        \toprule
        \multirow{2}{*}{\textbf{Models}} & \multicolumn{6}{c}{\textbf{MalVis-B Multi-Class}} \\
        \cmidrule(lr){2-7}
                                         & \textbf{A}            & \textbf{F1}         & \textbf{P}          & \textbf{R}          & \textbf{MCC}        & \textbf{ROC-AUC}     \\
        \midrule
        \textbf{MNv2}                    & 61.71\%               & 61.86\%             & 62.16\%             & 61.71\%             & 57.47\%             & 89.36\%              \\
        \textbf{DN201}                   & \cellcolor{LightBlue}\textbf{66.29\%}               & \cellcolor{LightBlue}\textbf{66.06\%}             & \cellcolor{LightBlue}\textbf{65.98\%}            & \cellcolor{LightBlue}\textbf{66.29\%}            & \cellcolor{LightBlue}\textbf{62.57\%}             & \cellcolor{LightBlue}\textbf{91.24\%}              \\
        \textbf{RN50}                    & 65.00\%              & 64.81\%             & 64.77\%            & 65.00\%            & 61.12\%             & 90.58\%              \\
        \textbf{INC-V3}                  & 64.43\%               & 64.37\%             & 64.43\%             & 64.43\%             & 60.48\%             & 90.35\%              \\
        \textbf{VGG16}                   & 60.70\%               & 60.23\%             & 60.02\%             & 60.70\%             & 56.36\%             & 89.35\%              \\
        \bottomrule
    \end{tabular}
\end{table}

\subsection{Evaluation of MalVis-B Performance using Ensemble Models on a Balanced Multiclass Dataset}
To address the performance impact caused by the undersampling approach, we explored the application of various ensemble methods. The aim was to take advantage of the combined strengths of all CNN models, which enhanced both the models' performance and robustness. The ensemble methods implemented and evaluated include:
\begin{itemize}
  \item \textbf{Average Voting:} Combines predictions by averaging the probabilities of all CNN models.
  \item \textbf{Majority Voting:} Determines the final class by selecting the most predicted by individual models.
  \item \textbf{Weighted Voting:} Assigns different weights to CNN models based on their prediction accuracy. We preserve the ranking performance of the models and assign weights corresponding to their place in the ranking.
  \item \textbf{Min Confidence Voting:} Only consider a model's prediction when it meets the minimum required confidence level. In our implementation, a confidence threshold of 60\% was selected.
  \item \textbf{Soft Voting:} Uses the predicted class probabilities to decide the final output.
  \item \textbf{Median Voting:} Determines decisions by selecting the median of predicted class probabilities.
  \item \textbf{Rank-Based Voting:} Ranks predictions from models and aggregates ranks to select a class.
  \item \textbf{Stacking Ensemble:} Trains a new model to integrate the predictions of the base model and improve performance.
\end{itemize}

\begin{figure*}[p]
  \centering
  \includegraphics[height=0.93\textheight]{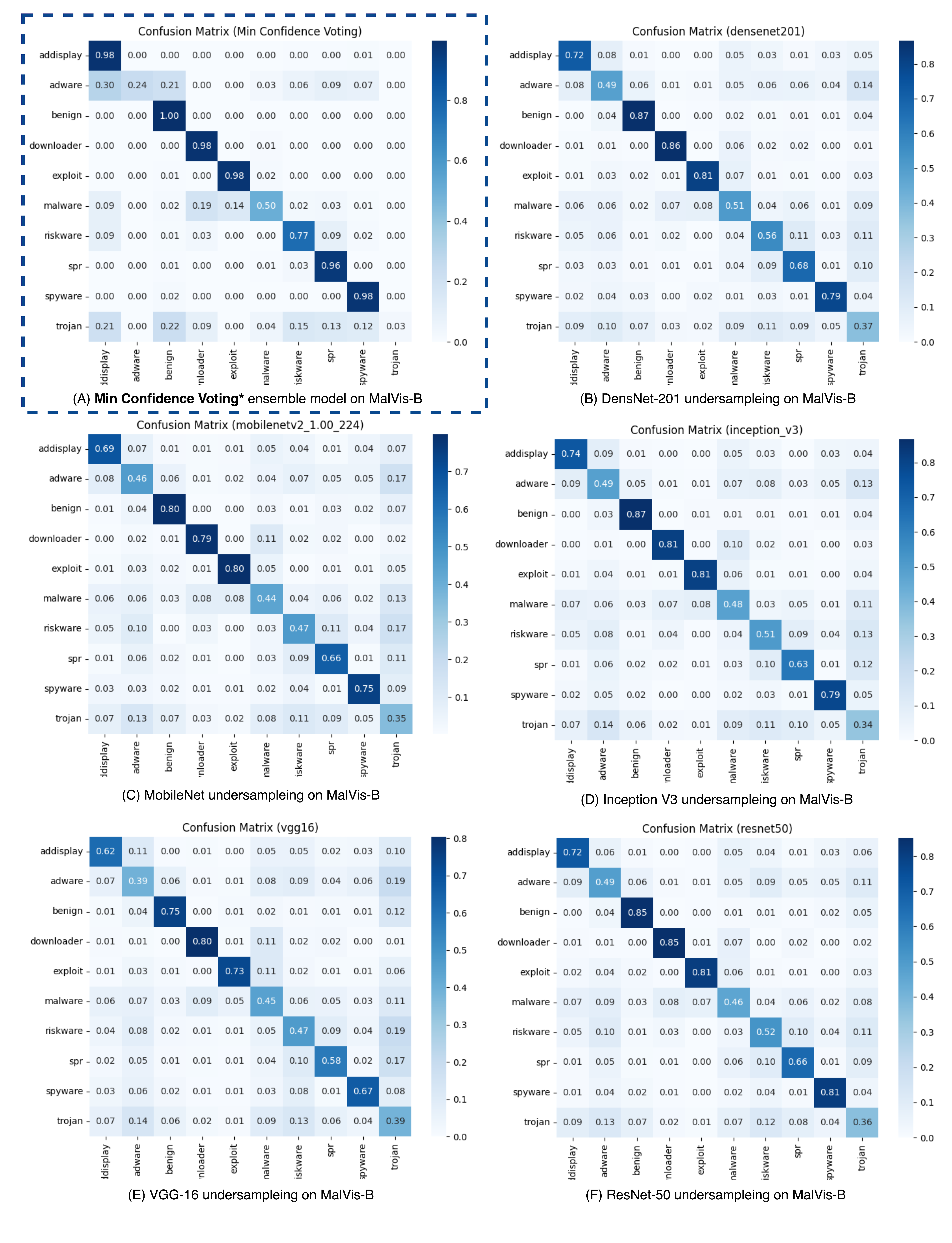}
  \caption{Confusion matrices for CNN models trained on a balanced multiclass MalVis dataset and the optimal ensemble method.}
  \label{fig:confusion matrix undersampleing}
\end{figure*}

In Table~\ref{tab:MalVis-B in multiclass}, the Min Confidence Voting ensemble achieved the highest performance across all evaluation metrics except for ROC-AUC. These results signify superior performance compared to the results in the unbalanced dataset shown in Table~\ref{tab:results-multi}. The confusion matrix in Fig.~\ref{fig:confusion matrix undersampleing} in box $\boxed{A}$ illustrates that the Min Confidence Voting ensemble demonstrated enhanced performance by producing a more pronounced diagonal shape. This indicates an improved ability to accurately detect the more challenging classes compared to the CNN models shown in boxes $\boxed{B}$ to $\boxed{F}$ after undersampling. Moreover, the Stacking ensemble achieved the highest ROC-AUC metric, attributable to its ability to integrate predictions from multiple models, thereby leveraging their strengths to improve overall performance in distinguishing different classes. 

These findings underline the effectiveness of ensemble methods, particularly Min Confidence Voting and Stacking, in handling multiclass classification challenges on the Large-MalVis dataset.
\begin{table}[ht]
\centering
    \normalsize
    \caption{Performance results of different ensemble methods on the Large-MalVis multiclass dataset after undersampling evaluation.}\label{tab:MalVis-B in multiclass}
    \begin{tabular}{@{}lccccc@{}}
        \toprule
        \multirow{2}{*}{\textbf{Ensemble Methods}} & \multicolumn{5}{c}{\textbf{MalVis-B Multi-Class}} \\
        \cmidrule(lr){2-6}
                                                  & \textbf{A}            & \textbf{F1}         & \textbf{P}          & \textbf{R}          & \textbf{ROC-AUC}     \\
        \midrule
        \textbf{Average Voting}                  & 66.25\%               & 65.15\%             & 65.36\%             & 66.25\%             & 81.49\%              \\
        \textbf{Majority Voting}                 & 63.15\%               & 61.40\%             & 63.70\%             & 63.15\%             & 79.79\%              \\
        \textbf{Weighted Voting}                 & 64.79\%               & 63.15\%             & 63.90\%             & 64.79\%             & 80.72\%              \\
        \textbf{Min Confidence Voting}           & \cellcolor{LightBlue}\textbf{88.65\%} & \cellcolor{LightBlue}\textbf{86.32\%} & \cellcolor{LightBlue}\textbf{89.02\%} & \cellcolor{LightBlue}\textbf{88.65\%} & 86.41\% \\
        \textbf{Soft Voting}                     & 66.25\%               & 65.15\%             & 65.36\%             & 66.25\%             & 81.49\%              \\
        \textbf{Median Voting}                   & 64.51\%               & 63.49\%             & 64.39\%             & 64.51\%             & 80.54\%              \\
        \textbf{Rank-Based Voting}               & 63.23\%               & 63.12\%             & 64.34\%             & 63.23\%             & 79.82\%              \\
        \textbf{Stacking Ensemble}               & 83.61\%               & 83.33\%             & 83.50\%             & 83.61\%             & \cellcolor{LightBlue}\textbf{90.99\%} \\
        \bottomrule
    \end{tabular}
\end{table}

\section{Conclusions}~\label{sec:conclusion}

This research establishes the critical importance of visualizing Android malware to safeguard user data and smartphone security. We introduced {MalVis}, the largest publicly available image-based dataset for Android malware, containing over 1.3 million samples. To complement this resource, we developed a novel visualization framework that transforms bytecode into RGB images by integrating entropy and N-gram encoding techniques. This method effectively captures encryption, compression, structural, and operational anomaly patterns within the malware. 

Through extensive evaluation, MalVis consistently outperformed existing visualization-based detection approaches, achieving 95.19\% accuracy, 90.81\% F1-score, 92.58\% precision, 89.10\% recall, 87.58\% Matthews Correlation Coefficient, and a 98.06\% ROC-AUC. Beyond its strong performance, MalVis delivers a conceptually innovative framework that links visual representations to the semantic characteristics of malware, enhancing interpretability and classification robustness. This dataset and framework provide a valuable foundation for advancing research in malware classification, adversarial resilience, and explainable threat detection.

\clearpage

\clearpage

\clearpage
\bibliographystyle{unsrt}  
\bibliography{REF}  

\end{document}